\documentclass[journal]{IEEEtran}

\usepackage{pgfplots}
\usepackage{verbatim}
\usepackage{multirow}
\usepackage{amsmath}
\usepackage{tabularx}
\usepackage{nomencl}
\usepackage{comment}

\makenomenclature

\usepackage{etoolbox}
\renewcommand\nomgroup[1]{%
  \item[\bfseries
  \ifstrequal{#1}{I}{Indices}{%
  \ifstrequal{#1}{V}{Variables}{%
  \ifstrequal{#1}{P}{Parameters}{%
  \ifstrequal{#1}{S}{Sets}}}}%
]}

\usepackage{pgf-pie}

\title{Impacts of Variable-Impedance-Based Power Flow Control on Renewable Energy Integration}
\author{Omid~Mirzapour,\IEEEmembership{Student Member,~IEEE}
        Mostafa Sahraei-Ardakani,~\IEEEmembership{Member,~IEEE}
        \thanks{The authors are with the Department of Electrical and Computer Engineering, University of Utah, Salt Lake City, UT 84112 USA (email: omid.mirzapour@utah.edu; mostafa.ardakani@utah.edu). This research was funded by the National Science Foundation CAREER award number 2146531.}
        }
\date{May 2020}

\pgfplotsset{compat=1.16}
\begin{document}

\maketitle

\begin{abstract}
The electric power grid has evolved significantly over the past two decades in response to climate change. Increased levels of renewable energy generation, as a prominent feature of this evolution, have led to new congestion patterns in the transmission network. The transmission system is originally designed for conventional energy sources, with predictable flow patterns. Insufficient transfer capability in congested transmission systems results in commitment of more expensive power plants and higher levels of renewable energy curtailment. One way to mitigate congestion is adoption of power flow control through variable-impedance flexible ac transmission system (FACTS) devices. In this paper the impacts of power flow control on generation cost, carbon emissions and renewable energy curtailment are studied under a wide range of scenarios, including generation mix from major US regional transmission organizations, and different load curves, representing seasonal variations. A two-stage stochastic unit commitment, including FACTS adjustment, is used to evaluate the impacts of FACTS devices on various types and penetration levels of renewable energy. The results show that FACTS installation effectively reduces generation cost, carbon emissions, and renewable energy curtailment. Location of renewable energy resources, peak-hour demand and the system's generation mix are among the influential factors.
\end{abstract}

\begin{IEEEkeywords}
Carbon emissions, flexible ac transmission systems (FACTS), power flow control,  renewable energy, solar energy, stochastic unit commitment, weather variability, wind energy.
\end{IEEEkeywords}

\nomenclature[V]{$PL$}{Real power flow through line}
\nomenclature[P]{$B$}{Transmission line susceptance}
\nomenclature[V]{$\theta^S$}{Voltage angle on sending bus}
\nomenclature[V]{$\theta^R$}{Voltage angle on receiving bus}
\nomenclature[P]{$B^{max}$}{Maximum susceptance for transmission line equipped with FACTS }
\nomenclature[P]{$B^{min}$}{Minimum susceptance for transmission line equipped with FACTS }
\nomenclature[V]{$F$}{Transmission line flow direction}
\nomenclature[P]{$P^W$}{Wind power generation}
\nomenclature[P]{$\nu$}{Wind speed}
\nomenclature[P]{$P_{rated}$}{Wind turbine rated power}
\nomenclature[P]{$\nu_{rated}$}{Wind turbine rated wind speed}
\nomenclature[P]{$\nu_{ci}$}{Wind turbine cut-in wind speed}
\nomenclature[P]{$\nu_{co}$}{Wind turbine cut-out wind speed}
\nomenclature[P]{$P^S$}{Solar power generation}
\nomenclature[P]{$I_O$}{PV array output current}
\nomenclature[P]{$I_{SC}$}{PV array short circuit current}
\nomenclature[P]{$V_{OC}$}{PV array open circuit voltage}
\nomenclature[I]{$g$}{Generator}
\nomenclature[P]{$G$}{Total number of generators}
\nomenclature[I]{$t$}{Time}
\nomenclature[P]{$T$}{Time horizon}
\nomenclature[P]{$c^{nl}_g$}{Minimum generation cost}
\nomenclature[V]{$u_{gt}$}{Generator up/down status}
\nomenclature[P]{$c_g^{su}$}{Generator start-up cost}
\nomenclature[V]{$v_{gt}$}{Generator start-up variable}
\nomenclature[P]{$c_g^{sd}$}{Generator shut-down cost}
\nomenclature[V]{$w_{gt}$}{Generator shut-down variable}
\nomenclature[I]{$k$}{Piece-wise linear cost function segment}
\nomenclature[P]{$K$}{Total number of segments in piece-wise linear cost function}
\nomenclature[P]{$c_{gk}^{seg}$}{Piece-wise linear generation cost}
\nomenclature[V]{$P_{gtk}^{seg}$}{Real power generated in the $k$th segment of generator}
\nomenclature[I]{$s$}{Scenario}
\nomenclature[P]{$S$}{Total number of scenarios}
\nomenclature[P]{$\pi_{st}$}{Scenario probability at time $t$}
\nomenclature[P]{$c_g^{UE}$}{Energy deployment cost}
\nomenclature[V]{$P_{gts}^{ru}$}{Real power ramp-up}
\nomenclature[V]{$P_{gts}^{rd}$}{Real power ramp-down}
\nomenclature[I]{$r$}{Renewable energy resource}
\nomenclature[P]{$R$}{Total number of renewable energy resources}
\nomenclature[P]{$c_r^{RC}$}{Renewable energy curtailment cost}
\nomenclature[V]{$P_{rts}^{RC}$}{renewable energy curtailment}
\nomenclature[V]{$x_l^f$}{FACTS allocation variable}
\nomenclature[P]{$c_h^{FACTS}$}{hourly investment cost of FACTS}
\nomenclature[P]{$c^{FACTS}$}{FACTS device investment cost}
\nomenclature[P]{$S_{FACTS}$}{FACTS device maximum compensation rating}
\nomenclature[P]{$PL^{max}$}{Transmission line thermal rating}
\nomenclature[P]{$S_{base}$}{MVA base of the system}
\nomenclature[V]{$P_{gt}$}{Generator real power generation}
\nomenclature[P]{$P_g^{max}$}{Generator upper generation limit}
\nomenclature[P]{$P_g^{min}$}{Generator lower generation limit}
\nomenclature[P]{$UT_g$}{Generator minimum up time}
\nomenclature[P]{$DT_g$}{Generator minimum down time}
\nomenclature[P]{$RU_g$}{Generator per-minute ramp-up rate}
\nomenclature[P]{$RD_g$}{Generator per-minute ramp-down rate}
\nomenclature[I]{$l$}{Transmission line}
\nomenclature[P]{$L$}{Total number of transmission lines}
\nomenclature[P]{$P^R_{rts}$}{Renewable energy generation}
\nomenclature[P]{$P^D_{bt}$}{Real power demand at bus $b$}
\nomenclature[S]{$NG_b$}{Set of generators located at bus $b$}
\nomenclature[S]{$NL_b^+$}{Set of transmission lines flowing into bus $b$}
\nomenclature[S]{$NL_B^-$}{Set of transmission lines flowing from bus $b$}
\nomenclature[S]{$NR_b$}{Set of renewable energy resources located at bus $b$}
\nomenclature[I]{$b$}{Bus}
\printnomenclature

\section{Introduction}

\IEEEPARstart{G}{lobal} warming has become a universal concern as its negative impacts on human livelihood is becoming increasingly more apparent. Carbon dioxide, making up for 81\% of greenhouse gases, is the main factor leading to global warming~\cite{GHG}. Worldwide, fossil fuel power plants, as the main source of electricity generation, produce 27\% of the total carbon dioxide~\cite{GHG}. To battle climate change, renewable energy resources have been introduced as cost-effective, emission-free alternatives to fossil fuel power plants. United States, as an instance, has increased its renewable energy supply by 50\% during the past decade~\cite{RE} and has an ambitious goal of reaching a carbon-free grid by 2035~\cite{BIDEN}. This is an initial step towards the goal of a carbon free US economy by 2050, where other polluting sectors, e.g., transportation, are either electrified or evolved to use other clean sources of energy, such as hydrogen~\cite{BIDEN2}.

In contrast with conventional fossil-fueled thermal units, renewable energy resources are intermittent in nature and more or less not dispatchable.  Transmission networks, conventionally designed to handle dispatchable generation, have faced difficulties handling the variability of renewable energy resources. This has led to renewable energy curtailment, due to transmission constraints. United States balancing authorities experienced an average of 1\% to 4\% of wind curtailment between 2007 and 2013~\cite{RC}. California Independent System Operator (CAISO) with 7,800 MW of wind and 15,000 MW of solar generation capacity installed as of 2022, has experienced large amounts of renewable energy curtailment mainly due to congestion~\cite{CAISO}. CAISO is seeking to cope with this challenge by adopting several roadmaps, including energy imbalance market expansion, demand response and electric vehicle charging coordination~\cite{CAISO_S}. Electric Reliability Council of Texas (ERCOT) with 25\% share of wind generation as of 2021, experienced an average of 8\% of wind curtailment, which peaked at 17\% in 2009~\cite{RC,ERCOT}. The curtailment was reduced to about 1.6\% in 2013 after ERCOT carried out transmission expansion. However, the curtailment doubled with steep growth of renewable energy generation projects in the southern areas~\cite{BTU}. Furthermore, transmission expansion is a rather costly solution for the congestion caused by renewable energy resources. Mid-continent Independent System Operator (MISO) with 28.9\% of wind energy penetration, adopted Dispatchable Intermittent Resource (DIR) protocol in 2011 to address the recurring wind curtailment problem~\cite{MISO,GTECH}. PJM interconnection experienced about 80,000 MWh of wind curtailment with the lost opportunity cost of \$3 million during September 2012~\cite{RC}. In order to increase efficiency and reduce the curtailment, PJM has changed its curtailment signaling and compensation process in 2013 and has lowered maximum wind curtailment from 8\% in 2014 to less than 4\% in 2019~\cite{PJM}. In this new process, PJM notifies the wind generator to lower its output to follow base-point signals prior to curtailment. If wind units do not follow the automatic base-point signals and provide sufficient data for accurate generation forecast they receive no compensation for the curtailment~\cite{RC}.

Beside the solutions mentioned above, several other approaches have been suggested to reduce renewable energy curtailment in the literature. Energy storage has been proposed in~\cite{barton2004energy} including  battery storage, pumped hydro storage and compressed air storage in order to manage load-generation balance and decrease renewable energy curtailment. While energy storage can potentially solve many of today's challenges, the rather high cost is still a major obstacle for its adoption. Demand response through flexible loads has been suggested as another approach to match the demand with generation, but this method is limited by scarcity of flexible loads~\cite{wu2014thermal}. Since the main reason for renewable energy curtailment is transmission system congestion, an alternative approach is to enhance the transfer capability of the existing network by exploiting transmission flexibilities. Such flexibility can be offered by phase shifters~\cite{momoh2001power}, transmission switching~\cite{fisher2008optimal,hedman2010co} or flexible ac transmission system (FACTS) devices~\cite{sahraei2015fast,sahraei2015day}. FACTS devices can control various properties of power system such as voltage phase and magnitude, shunt susceptance or line impedance. Variable-impedance FACTS devices can be effectively utilized to control power flow. In~\cite{sang2017stochastic} a stochastic unit commitment model is proposed to optimally adjust FACTS set-points as well as thermal units' generation to minimize wind curtailment. References ~\cite{sahraei2015transfer,sahraei2018merchant} propose a framework to implement series FACTS devices in market environments to increase the transfer capability of the transmission system. The interdependence of variable-impedance FACTS devices and transmission switching in power flow control is shown in~\cite{sang2017interdependence}. However, the optimal power flow through variable-impedance FACTS devices is mainly focused on minimizing the generation cost with respect to transmission system constraints. Therefore, in some cases using cheaper energy resources such as coal-fueled units results in increased carbon greenhouse gases, carbon dioxide above all. This outcome may overshadow the environmental merits emanating from installing variable-impedance FACTS devices.

This paper extends our previous study~\cite{9449793} on  impacts of variable impedance FACTS devices on carbon dioxide emissions, renewable energy curtailment, and generation cost through a stochastic unit commitment that co-optimizes FACTS set points and thermal generation dispatch, while accounting for the penetration of wind and solar units. Furthermore, influence of several factors including renewable energy penetration levels, wind and solar unit locations and dispatchable generation mix on emissions and curtailment are investigated with variable-impedance FACTS installed in the network. Simulations are carried out on a 24-bus modified RTS-96 system with the generation data from major regional transmission organizations in the US , including PJM, CAISO, MISO, ISO-NE and ERCOT, as well as original RTS-96 generation mix. The results show that FACTS installation can effectively reduce generation cost and renewable energy curtailment. However, in some cases, where the FACTS devices are installed in proximity of low-cost generation units with high emissions, they can adversely increase the emission. Additionally, the generation mix plays an important role in power flow control efficiency as FACTS devices have more impact on cases with larger share of high-emission units in the generation mix. The effectiveness of power flow control is better achieved during highest demand periods, where installing FACTS can effectively reduce generation cost as well as carbon emissions by allowing cheaper generation units, including renewables, to dispatch generated power through the transmission network. The renewable energy curtailment levels are also reduced during lightly loaded hours with implementation of power flow control.

The remainder of this paper is organized as follows. Section II introduces variable-impedance FACTS model as well as models for wind and solar units. It is followed by the stochastic unit commitment model for co-optimizing FACTS set point adjustment and dispatchable generation. Section III includes test system and renewable units as well as FACTS devices specifications for this study. Simulation results are presented in Section IV and finally, Section V concludes the paper.

\section{Model Formulation}
Variable-impedance control can be implemented using several FACTS technologies including thyristor-controlled series compensators (TCSC) and the recently introduced Smart Wire Grid device~\cite{kreikebaum2010smart}. TCSC devices have been effectively utilized to control power flow and transmission system losses in the grid~\cite{abdel2003power,fuerte2000thyristor,feng2001allocation}. Moreover,~\cite{sahraei2016computationally,sang2019effective} proposed efficient linear models for integrating FACTS devices into grid optimization models. Variable-impedance FACTS devices can manipulate line susceptance $B$  within a range  to increase the transfer capability. Therefore, the real power flow constraint can be written as follows for each line equipped with FACTS:
\begin{equation}
    PL = B(\theta^S-\theta^R). 
\end{equation}

It should be noted that the equation above is nonlinear since the previously constant line susceptance $B$ is now treated as a variable through impedance control. Based on the method proposed in ~\cite{7153580}, this equation can be rewritten into two linear constraints by predicting line flow direction, $F$:
\begin{equation}
    B^{min}F(\theta^S-\theta^R)+B^{max}(1-F)(\theta^S-\theta^R)\leq PL, 
\end{equation}
\begin{equation}
    B^{max}F(\theta^S-\theta^R)+B^{min}(1-F)(\theta^S-\theta^R)\geq PL.
\end{equation}

%


In order to compare the cost saving with FACTS investment cost, the capital cost of FACTS devices can be converted into an hourly figure. The total cost of TCSC FACTS devices can be calculated as follows~\cite{feng2001allocation,alhasawi2012techno}:
\begin{equation}
    c^{FACTS}= 0.0015S^2_{FACTS}-0.713S_{FACTS}+153.75,
\end{equation}
where $S_{FACTS}$ is the maximum compensation rating for TCSC FACTS device and can be calculated as follows:
\begin{equation}
    S_{FACTS}=\frac{(PL^{max}_l)^2}{S_{base}}.
\end{equation}

The investment cost can be converted to an hourly figure using the discount rate and the lifespan of the device as follows:
\begin{equation}
    c_h^{FACTS}=\frac{r(1+r)^n}{(1+r)^n-1}*\frac{c^{FACTS}}{8760}.
\end{equation}

Wind generation can be modeled by the wind turbine model described in ~\cite{lubosny2003wind,al2017accurate}. The wind energy is attained between cut-in and cut-out wind speeds and is proportional to the cubic wind speed. Solar generation depends on both intrinsic characteristics of photo-voltaic (PV) panels, which are usually reported by current-voltage(I-V) and power-voltage (P-V) charts as well as extrinsic irradiation conditions. Wind speed and solar radiation both change within continuous ranges, which create infinite scenarios and makes it impractical to optimize the dispatchable generation and FACTS set points for the continuous uncertainty space. In order to overcome this, a smaller number of scenarios are selected by choosing representative ranges for wind speed and solar radiation and creating discrete scenarios for solar and wind generation accordingly. Using scenarios for modeling uncertainty, the stochastic dispatchable generation and FACTS adjustment co-optimization model can be formulated as follows.


  
  \begin{subequations}
  \label{eq:SCUC}
  \begin{flalign}
  \label{eq:SCUC-first}
  \begin{split}
       &\min \sum_{g=1}^G\sum_{t=1}^T c_g^{nl}u_{gt}+c_g^{su}v_{gt}+c_g^{sd}w_{gt}+\sum_{g=1}^G
      \sum_{t=1}^T\sum_{k=1}^K c_{gk}^{seg}P_{gtk}^{seg}\\
      &+ \sum_{g=1}^G\sum_{t=1}^T\sum_{s=1}^S \pi_{st} c_g^{UE}(P_{gts}^{ru}+P_{gts}^{rd})
      + \sum_{r=1}^R\sum_{t=1}^T\sum_{s=1}^S \pi_{st} c_r^{RC}P_{rts}^{RC}\\
      &+ T\sum_{l=1}^L x_l^f c_h^{FACTS}
  \end{split} &&
  \end{flalign}
  \begin{flalign}
  \label{eq:scuc1}
      P_{gt} = \sum_{k=1}^K P_{gtk}^{seg} \qquad\forall g,t &&
    \end{flalign}
    \begin{flalign}
    \label{eq:scuc2}
      P_{gt}+P_{gts}^{ru}-P_{gts}^{rd}\leq P_g^{max}u_{gt} \qquad\forall g,t,s &&
    \end{flalign}
    \begin{flalign}
    \label{eq:scuc3}
      P_{gt}+P^{ru}_{gts}-P_{gts}^{rd}\geq P_g^{min}u_{gt} \qquad\forall g,t,s &&
      \end{flalign}
      \begin{flalign}
      \label{eq:scuc4}
      v_{gt}-w_{gt} = u_{gt}-u_{gt-1} \qquad\forall g,t &&
      \end{flalign}
      \begin{flalign}
      \label{eq:scuc5}
      v_{gt}+w_{gt} \leq 1 \qquad\forall g,t &&
      \end{flalign}
      \begin{flalign}
      \label{eq:scuc6}
      \sum_{\tau=t-UT_g-1}^t v_{g\tau}\leq u_{gt} \qquad\forall g,t &&
      \end{flalign}
      \begin{flalign}
      \label{eq:scuc7}
      \sum_{\tau=t-DT_g-1}^t w_{g\tau}\leq 1-u_{gt} \qquad\forall g,t &&
      \end{flalign}
      \begin{flalign}
      \label{eq:scuc8}
      P_{gt}-P_{gt-1} \leq 60RU_gu_{gt-1}+10RUv_{gt} \hspace{0.2cm}\forall g,t\geq2 &&   
      \end{flalign}
      \begin{flalign}
      \label{eq:scuc9}
      P_{gt-1}-P_{gt} \leq 60RD_gu_{gt}+10RD_gw_{gt} \qquad\forall g,t\geq2&&
      \end{flalign}
      \begin{flalign}
      \label{eq:scuc10}
      0\leq P_{gts}^{ru}\leq 10RU_g \qquad\forall g,t,s&&
      \end{flalign}
      \begin{flalign}
      \label{eq:scuc11}
      0\leq P_{gts}^{rd}\leq 10RD_g \qquad\forall g,t,s&&    
      \end{flalign}
      \begin{flalign}
      \label{eq:scuc12}
      -PL^{max}\leq PL\leq PL^{max} \qquad\forall l,t,s&&
      \end{flalign}
      \begin{flalign}
      \label{eq:scuc13}
      \sum_{l=1}^L x_l^f \leq N_{FACTS}&&
      \end{flalign}
      \begin{flalign}
      \label{eq:scuc14}
      \begin{split}
       &x_l^f(F_lB_l^{min}+(1-F_l)B_l^{max})(\theta_{lts}^S-\theta^R_{lts})\hfill\\
       &+((1-x_l^f)B_l (\theta_{lts}^S-\theta^R_{lts})\leq PL_{lts}\qquad\forall l,t,s 
      \end{split}&&
      \end{flalign}
      \begin{flalign}
      \label{eq:scuc15}
      \begin{split}
        &x_l^f(F_lB_l^{max}+(1-F_l)B_l^{min})(\theta_{lts}^S-\theta^R_{lts})\\
        &+((1-x_l^f)B_l (\theta_{lts}^S-\theta^R_{lts})\geq PL_{lts}\qquad\forall l,t,s
      \end{split}&&
     \end{flalign}
    \begin{flalign}
    \label{eq:SCUC-last}
     \begin{split}
          &\sum_{g\in NG_b}(P_{gt}+P_{gts}^{ru}-P_{gts}^{rd})+\sum_{r\in NR_b}(P_{rts}^R-P^{RC}_{rts})\\
          &+\sum_{l\in NL_b^+}PL_{lts}-\sum_{l\in NL_b^-}PL_{lts}= P_{bt}^D\qquad \forall b,t,s
      \end{split} &&  
      \end{flalign}
       \end{subequations}
    
    The model seeks to minimize total generation cost (\ref{eq:SCUC-first}) while considering generator capacity constraints (\ref{eq:scuc1})-(\ref{eq:scuc3}), start-up and shut-down constraints (\ref{eq:scuc4})-(\ref{eq:scuc7}), and ramping constraints (\ref{eq:scuc8})-(\ref{eq:scuc11}). Line maximum flow constraint is given in (\ref{eq:scuc12}) and line flow equation in presence of variable impedance devices is presented in (\ref{eq:scuc14})-(\ref{eq:scuc15}). The nodal power balance equation is given in (\ref{eq:SCUC-last}). This model can be considered from two different viewpoints. If $x_l^f$ is taken as a decision variable the model describes a FACTS allocation problem which is a mixed-integer non-linear program (MINLP) due to the existence of the products of two decision variables. The NLP problem is computationally intensive; therefore, it needs to be linearized using big M transformation as in~\cite{sahraei2015fast}. However, in this study, as we intend to evaluate the impact of FACTS device locations on system properties, the FACTS devices are allocated to candidate lines, which are chosen based on engineering judgment. Therefore, $x_l^f$ is treated as a parameter and the formulation is a mixed-integer linear program (MILP) and can be solved in reasonable time with existing commercial optimization packages, such as CPLEX and Gurobi.
    
    \section{Test System Specifications}
    The studies are carried out on a modified single-area RTS-96 system with 24 buses~\cite{RTS}. 480 MW of load on buses 14, 15, 19, 20 are shifted to bus 13 and then loads on every bus in the system is increased by 5\% yielding total electricity demand of 59.660 GWh daily, considering the load curve data. In order to create congestion in the ratings of lines A25-1 and A25-2 are reduced to 175 MW and ratings of A21 and A22 are reduced to 220 MW. Three pairs of candidate buses (4,5) as representative for buses close to demand, (17,18) as buses close to low-cost energy resources and (3,24) as  typical buses in the system, are considered for renewable energy resources. Three candidate lines for FACTS device allocation are considered based on engineering judgement. Equipping highly utilized lines, lines with large capacity and lines with large reactance with FACTS devices has been shown to be most effective in~\cite{sang2017stochastic}. A21 and A25-1 lines are taken as highly utilized lines and A26 is considered as a large-capacity line for FACTS allocation. 
    
    Two wind farms with rated wind speed of 14 m/s and cut-in wind speed of 4 m/s and cut-out wind speed of 25 m/s as well as two solar farms are considered based on data obtained from national renewable energy laboratory (NREL)~\cite{WD,SO}.

In this paper, thyristor-controlled series compensators (TCSC) are used to control power flow in transmission lines. This type of FACTS device operates in both capacitive and inductive modes. In capacitive mode, the TCSC increases suceptance of transmission line, which results in more active power flow through the line while in inductive mode by reducing line susceptance, TCSC re-routes power flow through other transmission lines. The maximum adjustment range of -80\% to +40\% is considered for TCSC device as in~\cite{gerbex2001optimal}.

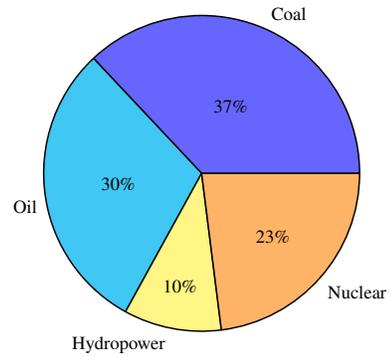
\begin{figure}[tbh!]
\centering
\scalebox{0.7}{\begin{tikzpicture}
\pie{37/Coal, 30/Oil, 10/Hydropower, 23/Nuclear}
\end{tikzpicture}}
\caption{RTS-96 Generation mix}
\label{fig-1}
\end{figure}

In order to evaluate carbon emissions by electricity generation, first the generation mix for the RTS-96 system needs to be known. The generation mix for the mentioned system is shown in figure\ref{fig-1}. Coal is an inexpensive and abundant source of energy.  Making 37\% of the generation in RTS-96, coal is the second largest source of energy in the United States and still the main source of carbon dioxide emissions in the power grid. Coal reserves are mainly available in four different types. The largest portion of coal resources is lignite, which has the lowest level of energy. Sub-bituminous coal with higher level of stored energy is the second prevalent type of coal. Bituminous, also known as soft coal, has the second density of thermal energy in coal types. Finally, anthracite is the rarest type of coal, although it has the highest level of stored energy. coal combustion produces other greenhouse gases such as sulfur dioxide ($SO_2$) and nitrogen oxides ($NO_x$) as well as carbon dioxide. Furthermore, coal mining process is a source of methane emission ($CH_4$) itself. The coal industry has adopted several methods to reduce emissions from coal-fired electricity generation, including desulfurization and carbon capture equipment development~\cite{Coal}. 

Heavy oil fuel with the smallest share in the generation mix, produces similar levels of greenhouse gases and is considered as a polluting source of energy. Roughly, 70\% of oil-fired generators was constructed prior to 1980. Oil-fired plants are generally committed during times of peak demand. These units have low capacity factors, mainly due to the high price of oil. since oil-fired generators are used to meet peak demand in general, they are designed to have low capacity factors and higher heat rates. Some  plants are capable of switching between oil and natural gas~\cite{Oil}. They burn natural gas for supplying baseline demand and oil to meet peak demand. Natural gas has surpassed coal, and is currently the leading generation source in the US. Natural gas-fired combined cycle plants are currently the most popular technology to supply baseload demand in the US. Other types of natural gas-fired plants including combustion turbines and steam turbines are committed during higher demand periods. Natural gas emits less greenhouse gases compared to oil and coal and is a cleaner source of energy, although it still produces similar levels of greenhouse gases. Natural gas-fired plants have experienced an upward trend during recent years as the capacity factor for gas fired generation in US has increased from 43\% in 2011 to 56\% in 2016~\cite{NG}. 

Hydropower, the largest renewable energy resource in US until recent years, has been surpassed by wind generation in 2019~\cite{EIA_Hydro}. In 2020, hydroelectricity comprised 6.6\% of   total electricity generation in U.S. and 22\% of renewable energy generation~\cite{EIA_share}. Hydropower, unlike fossil fuels, is an emission-free, cheap source of energy. However, hydropower expansion is limited by the availability of water. Conventional hydroelectric plants includes run-of-the-river systems, where the energy is supplied by the force of river's current, and reservoir systems, where the water is accumulated behind a dam and released through a turbine to generate electricity. Reservoir systems can be further upgraded to pumped hydro-storage that can pump water to a higher elevation during times of lower electricity price and release the power during peak load, when the electricity prices are high~\cite{HP}.

Nuclear power plants have generated 20\% of annual electricity generation in US since 1990. Nuclear power plants produce heat by nuclear fission  to generate steam. The steam goes through a turbine and then cooled back into water in a cooling tower or the water is supplied from the ocean or river close to the facility. Like natural gas-fired plants, nuclear plants are used to supply baseline demand. Nuclear power plants produce no greenhouse emissions. However, the nuclear waste produced in the electricity generation process in this type of plants is a major environmental concern. The radioactive radiation from the waste can remain dangerous for human beings and the environment for thousands of years. Therefore, it needs to be disposed under special regulations~\cite{NC}. Average operational generation cost and carbon emission for different types of plants is provided in Table~\ref{table-4} based on the data from~\cite{RTS,egrid}.  

\begin{table}[!t]
    \caption{Average Carbon Emission and Generation Cost for Plant Types}
    \label{table-4}
    \centering
    \renewcommand{\arraystretch}{1.5}
    \begin{tabularx}{\columnwidth}{l X X}
        \hline\hline
         & Emission Rate (lb/MWh) & Generation Cost (\$/MWh)\\
        \hline
        Coal-fired & 2027 & 22\\
        Oil-fired  & 1671 & 121\\
        Gas-fired  & 1169 & 14\\
        Nuclear    & 0       & 2\\
        Hydropower & 0       & 0\\ 
        
        \hline\hline
    \end{tabularx}
\end{table}

\section{Simulation Results}
In order to evaluate the impact of FACTS devices on generation cost, renewable energy curtailment and carbon emission, the co-optimization model described in (\ref{eq:SCUC-first})-(\ref{eq:SCUC-last}) is implemented on a modified RTS-96 system with specifications in previous section with a 24-hour time horizon. Renewable energy resources based on their type and location and the topology of the grid can create different congestion levels and therefore different cost saving by FACTS installation. In order to study the impacts of FACTS devices, simulations were carried out under a wide range of scenarios.

\subsection{Base Case Without Renewable Energy Integration}
First, FACTS devices are installed in RTS-96 system without any renewable integration to evaluate the FACTS impact on dispatchable generation. The results provide a baseline for total generation cost and carbon emission. As results in Table~\ref{table-5} indicate, installing FACTS devices on line 21 incurs more cost savings compared to equipping lines 25 and 26 with FACTS devices. This, however, increases carbon emissions, which is mainly due to the fact that line 21 is adjacent to coal-fired units that are inexpensive but have the highest level of carbon emissions among fossil fuel-fired plants. The objective in the co-optimization model is centered around minimizing total generation cost. Therefore, the FACTS settings are adjusted to maximize coal-fired generation, which results in increased amount of carbon emissions. The impact of power flow control can be directly observed in reducing congestion rent. Congestion rent, the difference between load payment and generation cost, range 6 to 10 percent of energy billing \cite{junginger2004cost}. Alleviating congestion reduces congestion rent and through that consumer payment by reducing locational marginal prices at nodes close to congested lines.
\begin{table}[!h]
    \caption{Simulation Results for RTS-96 Without Renewable Energy Resources}
    \label{table-5}
    \centering
    \renewcommand{\arraystretch}{1.2}
    \begin{tabularx}{\columnwidth}{X X X X X}
        \hline\hline
         Number of FACTS & FACTS Location (Line) & Total Generation Cost(M\$) & Congestion Rent (M\$) & Carbon Emission (Mlb) \\
        \hline
        0 & N/A & 1.988 & 0.248 & 66.551\\
        1 & 21  & 1.714 & 0.195 & 67.351\\
        2 & 25,26 & 1.885 & 0.247 & 63.662\\
        3 & 21,25,26 & 1.659 & 0.186 & 64.398\\
        
        \hline\hline
    \end{tabularx}
\end{table}

\subsection{Wind and Solar Farm Spacial Distribution}

In order to study the impacts of FACTS devices in presence of renewable energy resources 400 MW wind and solar farms are located on different buses with 4891.71 MWh total daily solar generation and 6412.42 MWh total daily wind generation using solar and wind scenarios as well as generation factors. 24 simulations were carried out in total with variable-impedance FACTS devices located on different lines. The results for the case with wind and solar integration  are shown in Tables \ref{table-6} and \ref{table-7} respectively. The results show that renewable energy integration effectively reduces total generation cost and carbon emission, which is expected since renewable energy resources are free and zero-emission sources. Therefore, replacing a part of fossil fuel-fired generation with renewable energy resources improves both emissions and cost savings. Wind generation results in higher cost savings and emission reductions compared to solar generation due to the fact that wind generation is less intermittent than solar generation. Equipping candidate lines with FACTS devices reduces the total generation cost in all cases as a result of the flexibility in power flow that FACTS devices can offer. However, the emissions are increased in cases that line 21 is equipped with FACTS, due to its proximity to high-emission inexpensive coal-fired units. In some cases, renewable energy curtailment is increased by installing FACTS devices. Although renewable energy is free, the transmission cost incurred by the congested lines may make other inexpensive units more economic than renewable energy resources. Therefore, the FACTS devices may adjust the power flow in a way that transmission system can deliver more power from other sources and curtail more renewable energy. Renewable energy location is a substantial factor in  total generation cost and carbon emissions. placing renewable energy resources on buses 17 and 18 brings less cost saving and emission reduction compared to other buses and more renewable energy is spilled when renewable energy resources are placed on these buses. This is a result of higher congestion rates in the proximity of these buses due to the closeness to low-cost energy resources.

\begin{table}[tbh!]
    \caption{Simulation Results for RTS-96 With Wind Integration}
    \label{table-6}
    \centering
    \renewcommand{\arraystretch}{1.2}
    \begin{tabularx}{\columnwidth}{X X X X X X}
        \hline\hline
        Wind Farm Location (Bus) & Number of FACTS & FACTS Location (Line) & Total Generation Cost(M\$) & Carbon Emission (Mlb) & Wind Curtailment (MWh) \\
        \hline
        \multirow{4}{*}{3,24}&0&N/A     &1.613&54.549&1390.43\\
                             &1&21      &1.401&55.798&1560.95\\
                             &2&25,26   &1.536&52.347&1527.21\\
                             &3&21,25,26&1.332&53.208&1649.85\\
        \hline                     
        \multirow{4}{*}{4,5} &0&N/A     &1.525&54.391&447.45\\
                             &1&21      &1.327&55.018&366.81\\
                             &2&25,26   &1.447&51.576&454.07\\
                             &3&21,25,26&1.267&52.278&364.65\\
       \hline
       \multirow{4}{*}{17,18}&0&N/A     &1.885&62.946&600.49\\
                             &1&21      &1.645&64.170&627.26\\
                             &2&25,26   &1.797&60.819&452.96\\
                             &3&21,25,26&1.556&61.193&452.96\\                     
        \hline\hline
    \end{tabularx}
\end{table}

\begin{table}[tbh!]
    \caption{Simulation Results for RTS-96 With Solar Integration}
    \label{table-7}
    \centering
    \renewcommand{\arraystretch}{1.2}
    \begin{tabularx}{\columnwidth}{X X X X X X}
        \hline\hline
        Solar Farm Location (Bus) & Number of FACTS & FACTS Location (Line) & Total Generation Cost(M\$) & Carbon Emission (Mlb) & Solar Curtailment (MWh) \\
        \hline
        \multirow{4}{*}{3,24}&0&N/A     &1.697&59.463&1434.28\\
                             &1&21      &1.453&60.677&1632.01\\
                             &2&25,26   &1.614&56.970&1460.31\\
                             &3&21,25,26&1.396&58.216&1697.13\\
        \hline                     
        \multirow{4}{*}{4,5} &0&N/A     &1.594&58.667&512.59\\
                             &1&21      &1.384&59.297&508.03\\
                             &2&25,26   &1.508&55.471&449.94\\
                             &3&21,25,26&1.311&56.397&453.73\\
       \hline
       \multirow{4}{*}{17,18}&0&N/A     &1.914&64.262&229.47\\
                             &1&21      &1.651&65.329&210.80\\
                             &2&25,26   &1.826&61.454&383.44\\
                             &3&21,25,26&1.579&62.396&383.44\\                     
        \hline\hline
    \end{tabularx}
\end{table}

\subsection{Renewable Energy Penetration Level}
\begin{figure}[tbh!]
\centering
\includegraphics[width=\columnwidth,keepaspectratio]{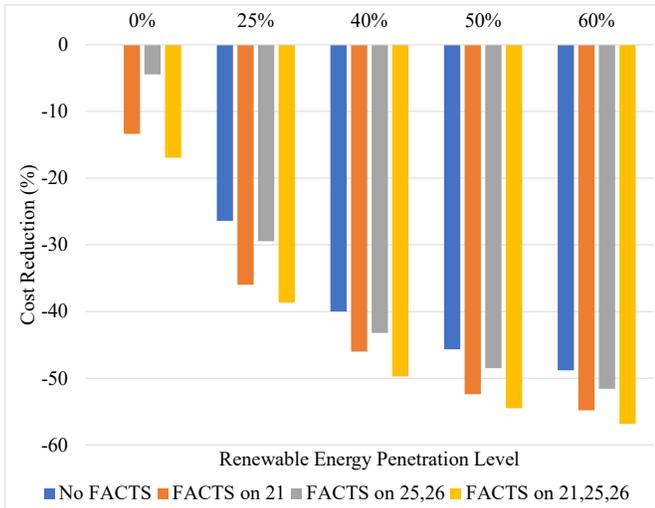}
\caption{Cost Saving for Renewable Energy Penetration Levels}
\label{fig-2}
\end{figure}

\begin{figure}[tbh!]
\centering
\includegraphics[width=\linewidth,keepaspectratio]{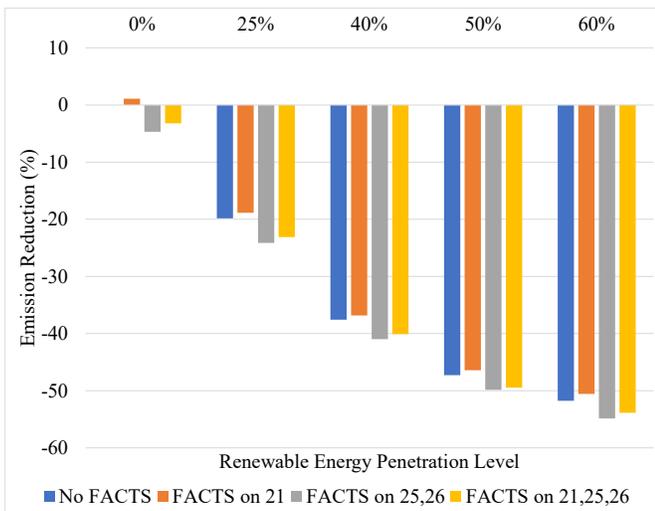}
\caption{Emission Reduction for Renewable Energy Penetration Levels}
\label{fig-3}
\end{figure}

\begin{figure}[tbh!]
\centering
\includegraphics[width=\linewidth,keepaspectratio]{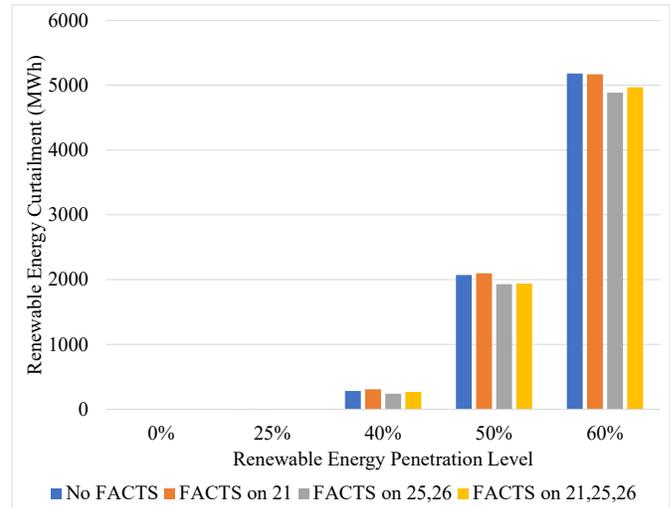}
\caption{Renewable Energy Curtailment for Renewable Energy Penetration Levels}
\label{fig-4}
\end{figure}

 In order to study the impact of FACTS devices under different penetration levels of renewable energy resources, wind and solar farms are distributed over all candidate buses with the same mixture of wind and solar capacity on each bus. Then, wind and solar capacities are increased in increments of 100 MW, each. Cost savings, emission reductions and renewable energy curtailments for different renewable energy penetration levels are shown in Fig. \ref{fig-2},\ref{fig-3},\ref{fig-4}. Higher levels of renewable energy penetration can reduce generation cost up to 34\%. the cost savings by renewable energy integration can be increased up to 46\% by installing FACTS devices on selected lines. Furthermore, renewable energy resources reduce carbon emissions by 29\% at maximum penetration level of 65\% which can be further increased to 32\% by implementing power flow control on lines 25,26 as shown in figure\ref{fig-3}. However, FACTS devices do not necessarily decrease renewable energy curtailment as explained before. This is shown in Fig. \ref{fig-4} for different renewable energy penetration levels.

\subsection{Load Curve}

\begin{figure}[tbh!]
\centering
\includegraphics[width=\linewidth,keepaspectratio]{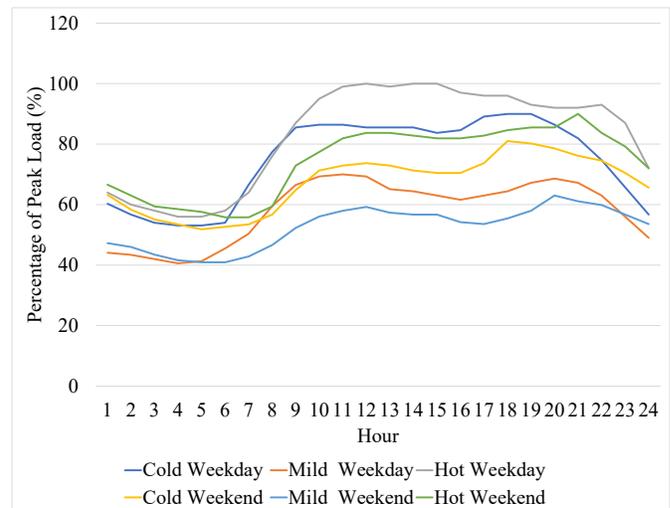}
\caption{Representative Load curves}
\label{loadcurve}
\end{figure}

\begin{figure}[tbh!]
\centering
\includegraphics[width=\linewidth,keepaspectratio]{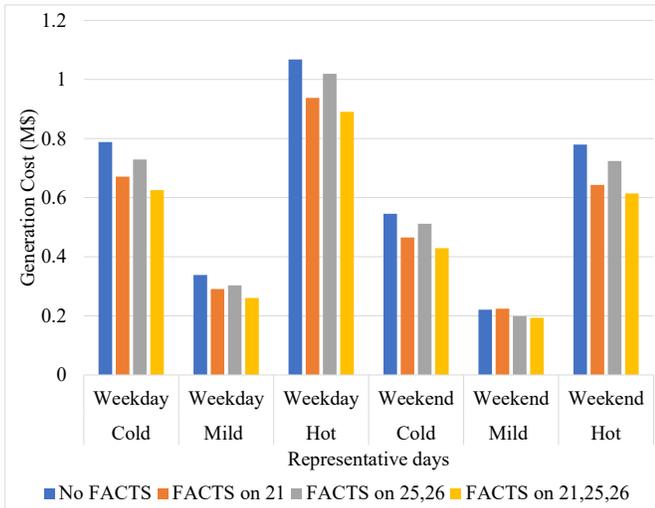}
\caption{Generation Cost for Representative Days}
\label{Cost_LC}
\end{figure}

\begin{figure}[tbh!]
\centering
\includegraphics[width=\linewidth,keepaspectratio]{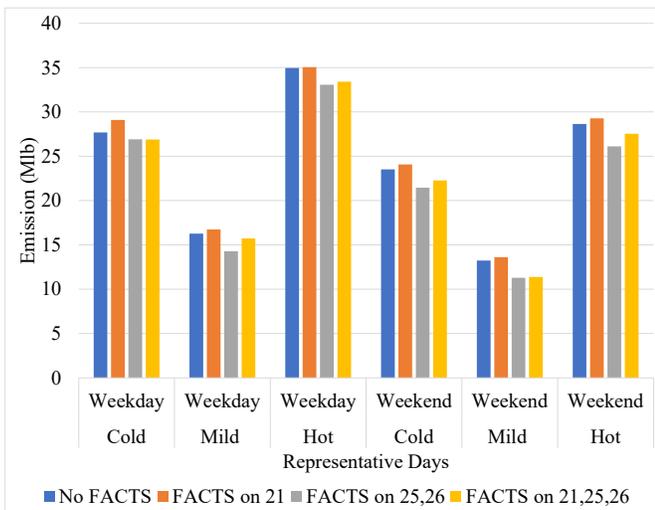}
\caption{Carbon Emission for Representative Days}
\label{Emission_LC}
\end{figure}

\begin{figure}[tbh!]
\centering
\includegraphics[width=\linewidth,keepaspectratio]{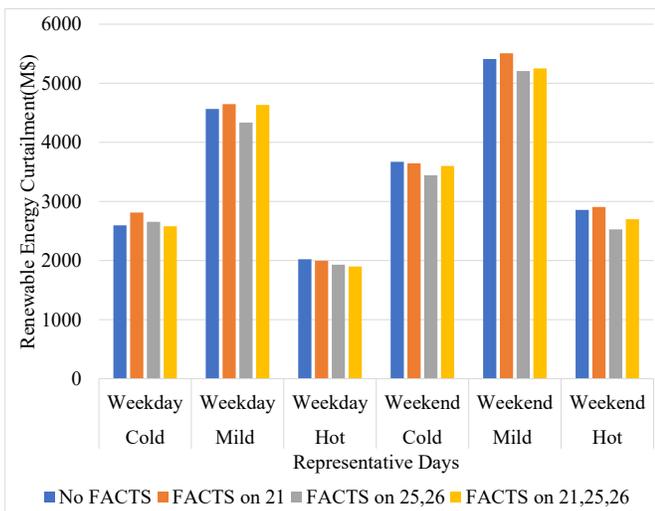}
\caption{Renewable Energy Curtailment for Representative Days}
\label{Curtailment_LC}
\end{figure}

With increased levels of renewable energy integration during the past the decade, the net demand curve is more weather driven and seasonal variation in temperature can be seen more distinctively than before \cite{staffell2018increasing}. To study the impact of impedance control under different load curves, 100 MW wind and solar units are placed at each candidate bus for renewables. Six representative load curves are considered for weekday and weekend during mild, cold and hot seasons as shown in Fig. \ref{loadcurve}. The distinct difference between weekend and weekday demand and number of peaking hours during hot and cold seasons are used to better study the effectiveness of variable impedance on generation cost, carbon emissions and renewable energy curtailment. As the Fig. \ref{loadcurve} shows, hot weekdays during summer incur the highest level of electricity demand and peaking hours due to electrified cooling systems' consumption. Cold seasons have a similar pattern with an average of 5\% lower demand compared to hot seasons. Finally the mild seasons have the lowest demand and a flat demand curve with a 10\% variation between the peak and baseline demand. The impact of impedance control on generation cost, carbon emissions and renewable energy curtailment is shown in Fig. \ref{Cost_LC},\ref{Emission_LC},\ref{Curtailment_LC}, respectively. Generation cost and carbon emissions are highest during hot season weekdays, where demand is at its highest level. Highest level of cost saving, incurred by impedance control with FACTS devices, is achieved during these days with 0.17 M\$ (16.6\%) cost saving. This is due to the fact that transmission system congestion appears during higher levels of demand and therefore, the impact of flexibility provided by FACTS devices is most considerable under this condition. However, the largest reduction in emissions is not necessarily achieved during the peak of demand, because with the congestion relieved, cheaper generating units with higher emission rates, including coal-fired units generate more power. Fig. \ref{Curtailment_LC} shows that highest level of renewable energy curtailment happens during mild weekends, when the electricity demand is at its lowest level. FACTS implementation helps reducing renewable energy curtailment by up to 204 MWh through enhancement of transfer capability.

\subsection{Generation Mix}

\begin{figure}[tbh!]
\centering
\includegraphics[width=\linewidth,keepaspectratio]{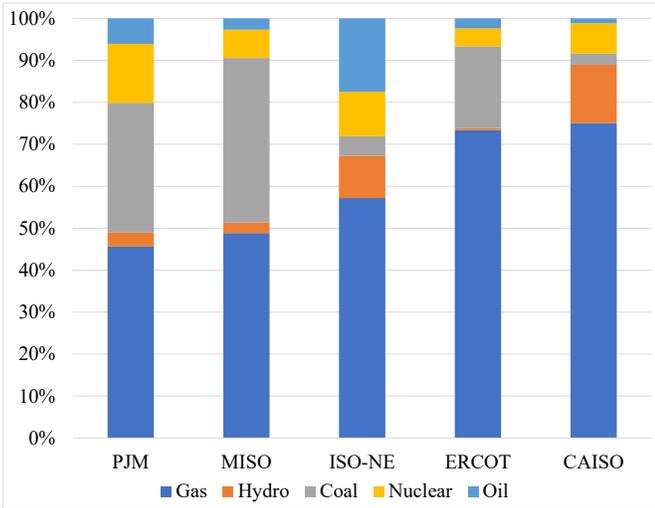}
\caption{Dispatchable Generation Mix for Major ISOs in the US }
\label{genmix}
\end{figure}

In order to study the impacts of FACTS with different generation mixes, generation capacity mixes for major ISOs in the US, including PJM, MISO, CAISO, ERCOT and ISO-NE have been implemented on the RTS system. Renewable penetration is simulated through 100 MW solar and wind farms placed at each candidate bus. Figure \ref{genmix} shows the generation mix for each ISO. Note that simulation studies in this section assume the same level of load for all the ISOs, using RTS load curves, and scales the supply to match the load as well as the ISO's generation mix.

PJM incorporates natural gas-fired units as the largest share of generation units with a total share of 45\%. Second largest share of PJM generation mix is coal-fired generation with 35\% of the total capacity. Renewable energy integration can reduce costs up to 28\%, which is further reduced to 30\% by power flow control implementation. Compared to original RTS-96 generation mix, impact of renewable energy integration and FACTS installation on cost saving is less for PJM generation mix. This due to the larger share of inexpensive generation sources in PJM. Due to the same reason, emission reduction by FACTS devices is less for PJM interconnection as the emission is reduced by 34\% through renewable integration and further reduced up to 36\% by equipping transmission lines with FACTS devices.    
MISO has the highest share of coal-fired units among major ISOs, with 39\% share of total generation capacity after 48\% share of natural gas units. As a result, it has the second highest greenhouse gas emissions after ERCOT. Equipping MISO's transmission system with FACTS, will reduce generation cost by 0.013 M\$. However, the emission is not reduced considerably, since the cheap resources in MISO are mostly coal-fired units, which result in increased emission.
ISO-NE has the largest capacity of oil-fired units (17\%) among other ISOs, which are expensive units with considerable level of carbon emission. Installing FACTS devices in this system will result in highest level of cost savings (6.8\%), as it eliminates the need for expensive oil-fired units. Both emission and renewable energy curtailment are reduced effectively by the flexibility provided in the grid. The emissions are reduced by 0.193 Mlb, while renewable energy curtailment is reduced by 96.4 Mwh.
ERCOT has the largest share of coal-fired units (19\%) after MISO and PJM. Therefore, the impact of power flow implementation on cost, emissions and renewable energy curtailment is very similar to MISO. FACTS devices can reduce generation cost by 0.011 M\$, while there is no considerable decline in emissions and renewable energy curtailment.



\begin{figure}[tbh!]
\centering
\includegraphics[width=\linewidth,keepaspectratio]{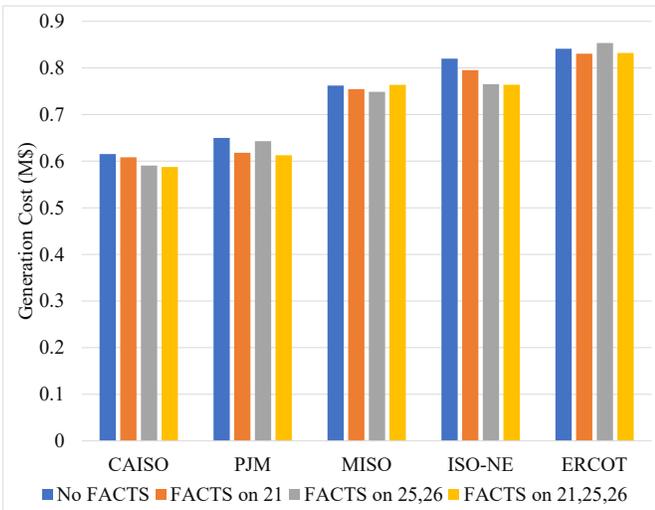}
\caption{Generation Cost for Major ISOs Equipped with FACTS}
\label{Cost_GM}
\end{figure}
\begin{figure}[tbh!]
\centering
\includegraphics[width=\linewidth,keepaspectratio]{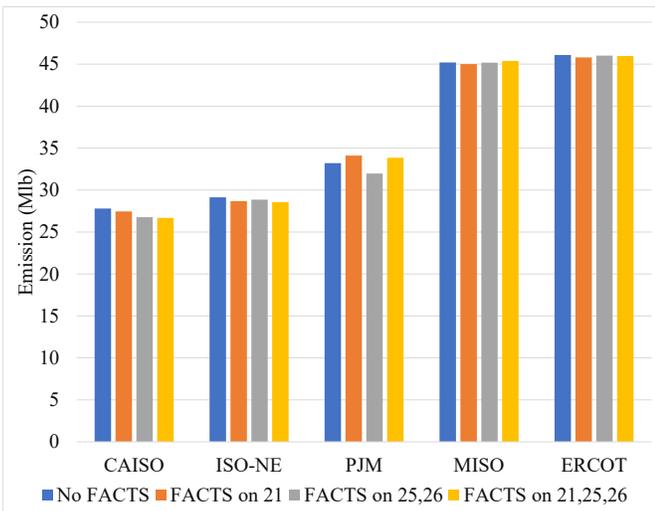}
\caption{Carbon Emission for Major ISOs Equipped with FACTS}
\label{Emission_GM}
\end{figure}
\begin{figure}[tbh!]
\centering
\includegraphics[width=\linewidth,keepaspectratio]{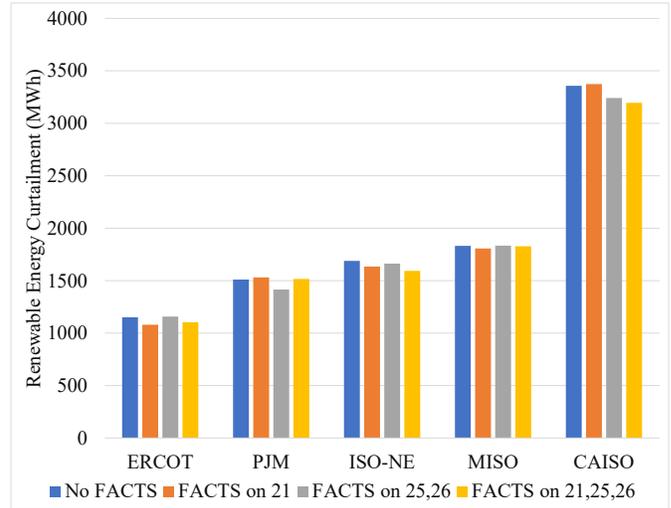}
\caption{Renewable Energy Curtailment for Major ISOs Equipped with FACTS}
\label{Curtailment_GM}
\end{figure}

CAISO has the largest share of renewable energy in form of hydropower. Natural gas-fired units make 75\% of CAISO's generation mix and hydropower with 14\% of total generation capacity stands in second place. Generation mix for CAISO is shown in Fig. \ref{genmix}. Due to large share of renewable energy and inexpensive gas-fired units, CAISO has already reduced its generation cost considerably. Therefore, equipping transmission lines with FACTS devices has the least impact on generation cost and carbon emissions in CAISO. 21\% cost saving by renewable energy integration is increased to 24\% with FACTS installation and emission reduced by 2\%  through controlling power flows, since the generation mix is already composed of low-emission inexpensive generation resources. Cost savings, carbon emission reductions and renewable energy curtailments are shown in Fig. \ref{Cost_GM},\ref{Emission_GM},\ref{Curtailment_GM} respectively.

\section{Conclusions}
Impacts of implementing power flow control on generation cost, carbon emissions and renewable energy integration were studied through a co-optimization model that adjusts FACTS device set-points alongside generation dispatch. The optimization model was implemented on a modified RTS-96 network to verify the impacts of variable impedance FACTS devices on the grid. The results show that installing FACTS devices reduces total generation cost in all cases. However, the cost savings are higher when the FACTS is installed close to inexpensive generation resources. Carbon emissions are reduced in most cases by equipping the transmission system with FACTS devices. However, when the host line is adjacent to high-emission units, e.g., coal-fired power plants, FACTS devices can increase carbon emissions. Integrating renewable energy resources, such as wind and solar units, reduces both generation cost and carbon emission. Installing FACTS devices increases the cost savings and often reduces emissions, but may increase renewable energy curtailment in cases where renewable energy resources are located close to other inexpensive resources. This is due to higher congestion rates in areas with lower generation cost. Studying power flow control under different load curves show that cost savings and carbon emissions, incurred by optimally operating FACTS devices, is best seen during high demand periods, such as summer days and weekdays, when the congestion pattern is at its highest level and FACTS devices can effectively reduce congestion. The power flow controllers can also reduce renewable energy curtailment for lightly loaded networks by optimally adjusting line impedances to allow highest levels of renewable energy dispatch. FACTS impact on generation cost and emission rates is better felt when generation mix contains higher levels of costly and high-emission energy resources. For RTOs like CAISO that already have replaced conventional high-emission fossil fueled plants with lower emission natural gas generation or hydropower generation, the impact of power flow control on cost savings is less significant. Nevertheless, power flow control effectively reduces renewable energy curtailment in networks like CAISO by reducing congestion. 

\bibliographystyle{IEEEtran}
\bibliography{Ref.bib}
\end{document}